\documentclass[a4paper,10pt]{article}
\usepackage{pos}
\usepackage{graphicx}
\usepackage{subfigure}
\usepackage{tikz}
\hypersetup{colorlinks=true,citecolor= blue,citecolor= blue}
\usepackage{bbm}
\usepackage{diagbox}
\usepackage{rotating}
\usepackage{float}
\restylefloat{table}
\usepackage{multirow}
\usepackage{sidecap}
\sidecaptionvpos{figure}{t}

\usepackage{hhline}
\usepackage{url}
\newcommand{\etal}{\textit{et al}. }

\let\OLDthebibliography\thebibliography
\renewcommand\thebibliography[1]{
  \OLDthebibliography{#1}
  \setlength{\parskip}{0pt}
  \setlength{\itemsep}{0pt plus 0.3ex}
}

\title{ Single- \& double-strangeness hypernuclei up to  $ A=8 $ within  chiral effective field theory }

\author[]{Hoai Le}
\affiliation[]{IAS-4, IKP-3 and JHCP, Forschungszentrum J\"ulich, D-52428 J\"ulich, Germany}
\emailAdd{h.le@fz-juelich.de}

\abstract{ 
We investigate  $S=-1$ and $-2$ hypernuclei with $A=4-8$ employing the Jacobi-NCSM approach and in combination with baryon-baryon interactions derived within the frame work of chiral effective field theory. The employed  interactions are transformed using    the  similarity renormalization group (SRG)  so that the low- and high-momentum states are decoupled, and, thereby,
 convergence of the binding energies with respect to model space can be significantly speeded up. Such an evolution is however   only approximately unitary when the so-called SRG induced higher-body forces are omitted.   We  first explore the impact of the SRG evolution on the $\Lambda$ separation energies $B_{\Lambda}$ in $A=3-5$ hypernuclei when only SRG-evolved two-body and when  both two- and three-body forces are included. For  the latter scenario, we thoroughly  study  predictions of 
  the two almost phase-equivalent  NLO13  and NLO19 YN  potentials for  $A=4-7$ hypernuclei.  The  NLO19 interaction  yields separation energies  that are comparable with experiment,  whereas NLO13 underestimates all the systems considered.  We further explore CSB splittings in the $A=7,8$ multiplets employing the two NLO YN potentials that include also the leading CSB potential in the $\Lambda$N channel, whose strength has been fitted to the presently established  CSB in $A=4$.  Finally, we report on  our recent study for $\Xi$ hypernuclei based on the $\Xi$N  interaction at NLO.    }

\FullConference{%
  *** The 14th International Conference on Hypernuclear and Strange Particle Physics - HYP2022***\\
  *** 27 June - 1 July 2022 ***\\
 }

\begin{document}
\maketitle

 \section{ {Introduction}}
 Experimental information on hyperon-nucleon (YN), hyperon-hyperon (YY) and $\Xi$ N scattering data is  indispensable input  for developing
 realistic YN and YY interactions \cite{Rijken:1999,Haidenbauer:2013,Haidenbauer:2020,Haidenbauer:2015zqb,Haidenbauer:2018gvg,Haidenbauer:2022prag}.  However, despite significant  progress over the last several decades, scattering data in the  strangeness $S=-1,-2$ sectors remain extremely  scarce and the situation is not expected to be substantially  improved in
 the near  future.   Yet, other important source of information  is  hypernucler spectroscopy \cite{Hashimoto:2006}, in which  not only  ground-state energies but also  energy level splittings in single- and double-strangness systems can be determined with  high accuracy.  This together with theoretical studies of hypernuclei, especially with microscopic approaches are  essential tools to explore the underlying baryon-baryon (BB) interactions and thereby to improve interactions  models.

In this study,  we  investigate  single-$\Lambda$ and $\Xi$  hypernuclear systems  with  $A=4 - 8$  based on the ``ab initio''  Jacobi no-core shell model (NCSM) \cite{Liebig:2016,Le:2020} and in combination with  BB  interactions derived within chiral effective field theory (EFT).   
In this  approach,  potentials are established in terms of an expansion  in powers  of low momenta  in accordance with an appropriate   power counting scheme.  The long-ranged parts of the interactions are fixed from chiral symmetry,  whereas unresolved short-distance dynamics are parameterized in terms of the so-called low energy constants (LECs), whose strengths should be    determined  via a fit to experiments.  In the non-strangeness sector, thanks to
 the wealth  of NN scattering data,  two-  and three-body  interactions  have been consistently  derived   up to high order,  e.g. $\mathrm{N}^4\mathrm{LO}$   \cite{Reinert:2018}  and $\mathrm{N^2LO}$ \cite{Maris:2021}, respectively, see also Ref.~\cite{Epelbaum:2022prag} for an overview on the current status of the chiral NN and 3N interactions. In the $S=-1,-2$ sectors, the situation is  however less satisfactory due to, as already said,  the lack of  the appropriate  scattering data. Despite of that, YN and YY potentials up to NLO  have 
 been derived \cite{Haidenbauer:2013,Haidenbauer:2020,Haidenbauer:2015zqb,Haidenbauer:2018gvg} and recently extended to $\mathrm{N^2LO}$ \cite{Haidenbauer:2022prag}.  Especially, at NLO,  by pursuing different strategies for fixing LECs,  two realizations of
the YN interaction,  NLO13 \cite{Haidenbauer:2013}  and NLO19 \cite{Haidenbauer:2020}, which yield
similarly good description of all the available  YN scattering data, could be established. However, as discussed in  \cite{Haidenbauer:2020}, NLO19 
is characterized by  a somewhat weaker $\Lambda \mathrm{N}-\Sigma$N transition potential, 
and as a result,   the two potentials yield slightly different predictions for 
$^4_{\Lambda}\mathrm{He}$ and for nuclear matter.  Also,  in  \cite{Le:2020}  we  have explored  effects  of NLO13 and NLO19
on light hypernuclei with   $A=4 - 7$,  but only qualitatively. Below we revisit this topic and take a closer look at the separation energies of these systems predicted by the  two potentials, and thereby,  discuss the possible effect of  chiral YNN forces.  In addition, we  explore  charge symmetry breaking (CSB) effects in the $A=7, 8$ multiplets,  employing the NLO13 and NLO19 interactions and with the inclusion of  the leading CSB potential derived in Ref.~\cite{Haidenbauer:2020csb}. The latter involves additional two LECs that have been determined via a fit to the  separation energies difference $\Delta B_{\Lambda}$ in the ground  and excited states  of  $^4_{\Lambda}$He and $^4_{\Lambda}$H, with $\Delta B_{\Lambda}(0^+)= 233 \pm 92$ keV and 
 $\Delta B_{\Lambda}(1^+)= -83 \pm 94$ \cite{Botta:2019has}, respectively. Finally, we  discuss  our recent work  \cite{Le:2021XN}, where  the possible existence of   $\Xi$ hypernuclei up to $A=7$  for  the  NLO $\Xi$N potential \cite{Haidenbauer:2018gvg}   has been explored.

\vspace{-0.5cm}

 
\section{ SRG  evolution for YN interactions and  impact on  $B_{\Lambda}$ in $A=3-5$ hypernuclei}
\begin{figure*}[htbp] 
       \vspace{-1cm}
      \begin{center}
         \subfigure[ ]{ \includegraphics[width=0.35\textwidth,trim={0.0cm 0.00cm 0.0cm 0.0cm},clip]{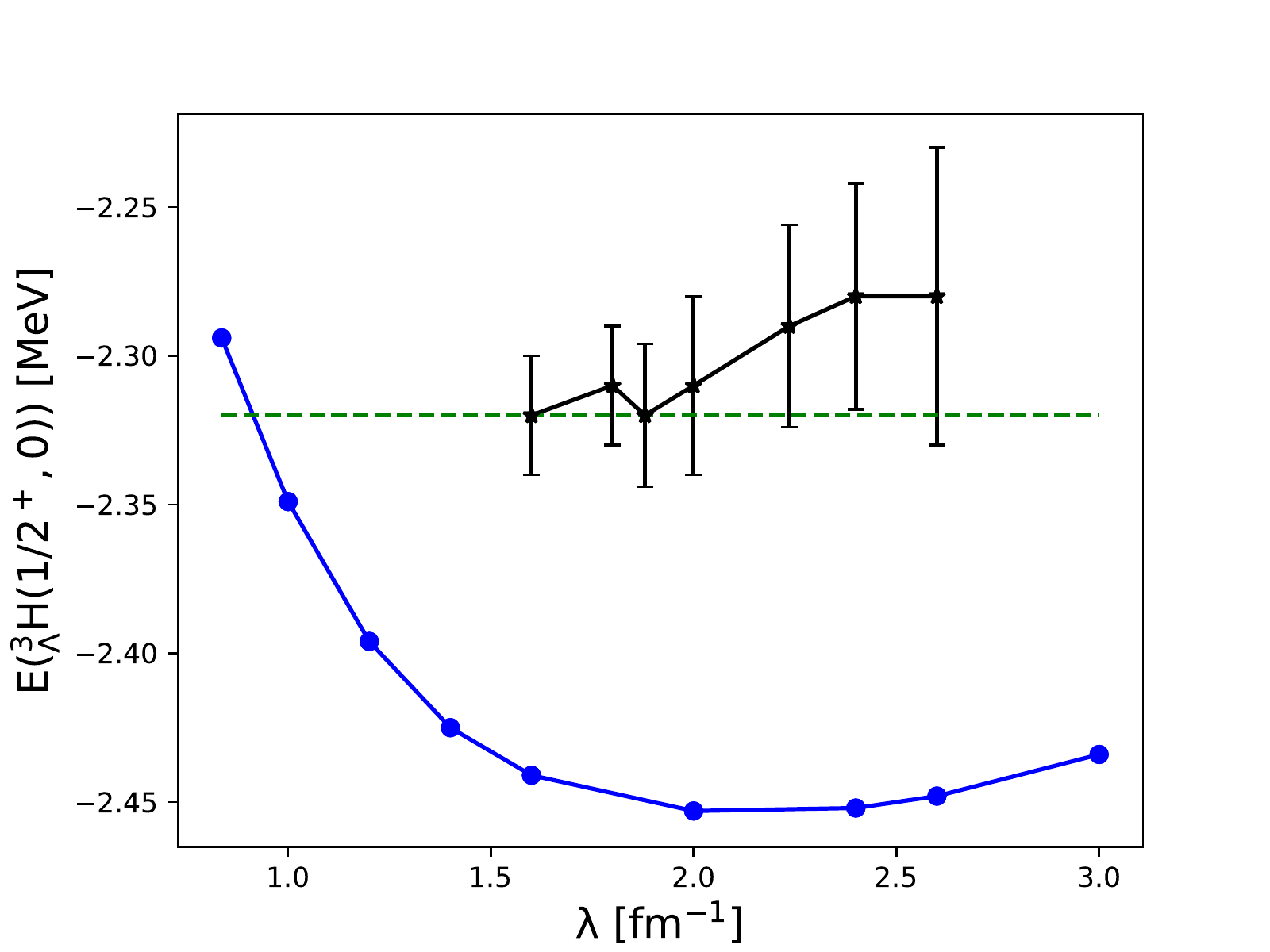}}
           \subfigure[ ]{ \includegraphics[width=0.35\textwidth,trim={0.0cm 0.00cm 0.0cm 0.0cm},clip]{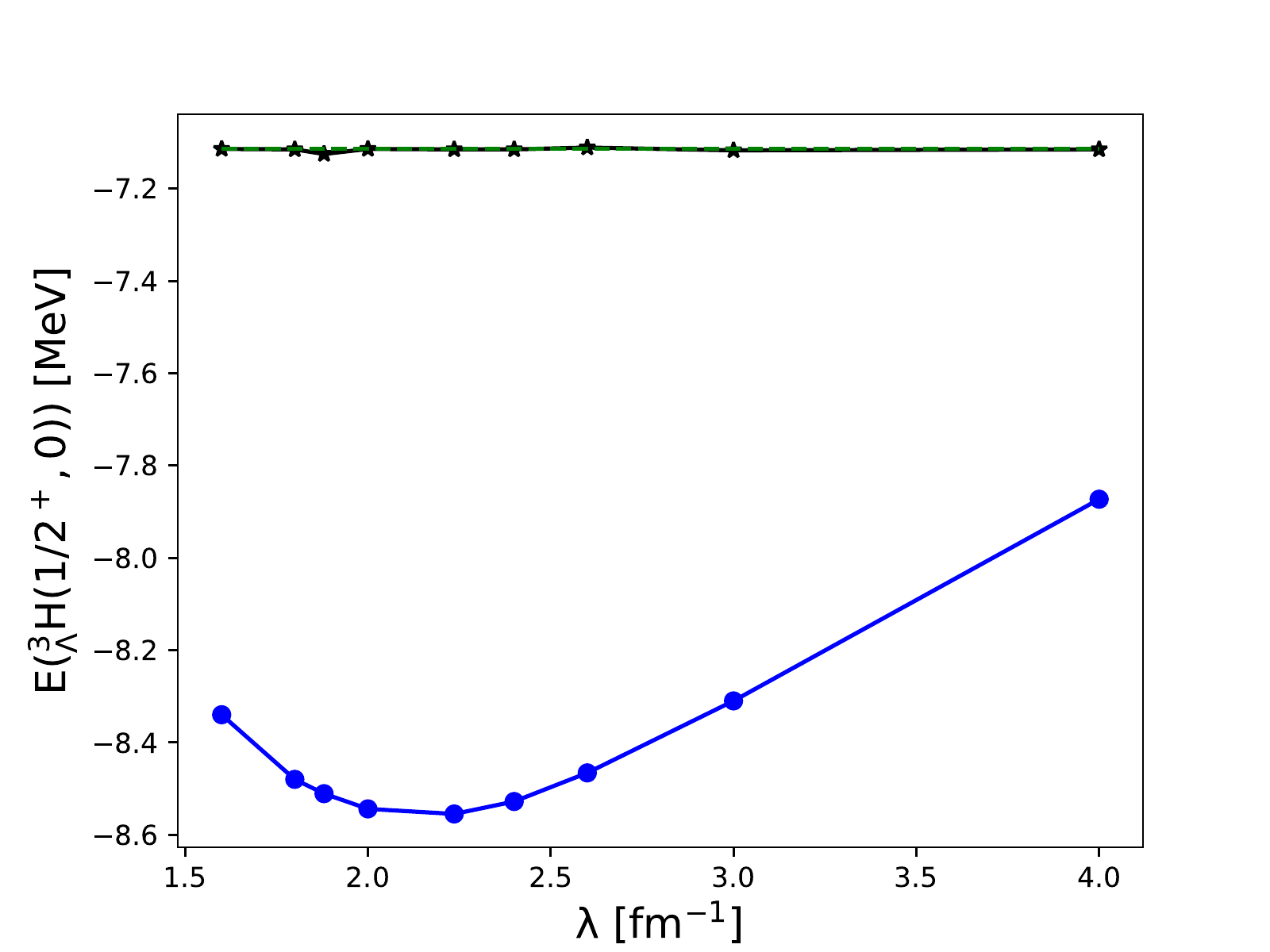}}\\[-8pt]
           \vspace{-0.135cm}
             \subfigure[ ]{ \includegraphics[width=0.35\textwidth,trim={0.0cm 0.00cm 0.0cm 0.0cm},clip]{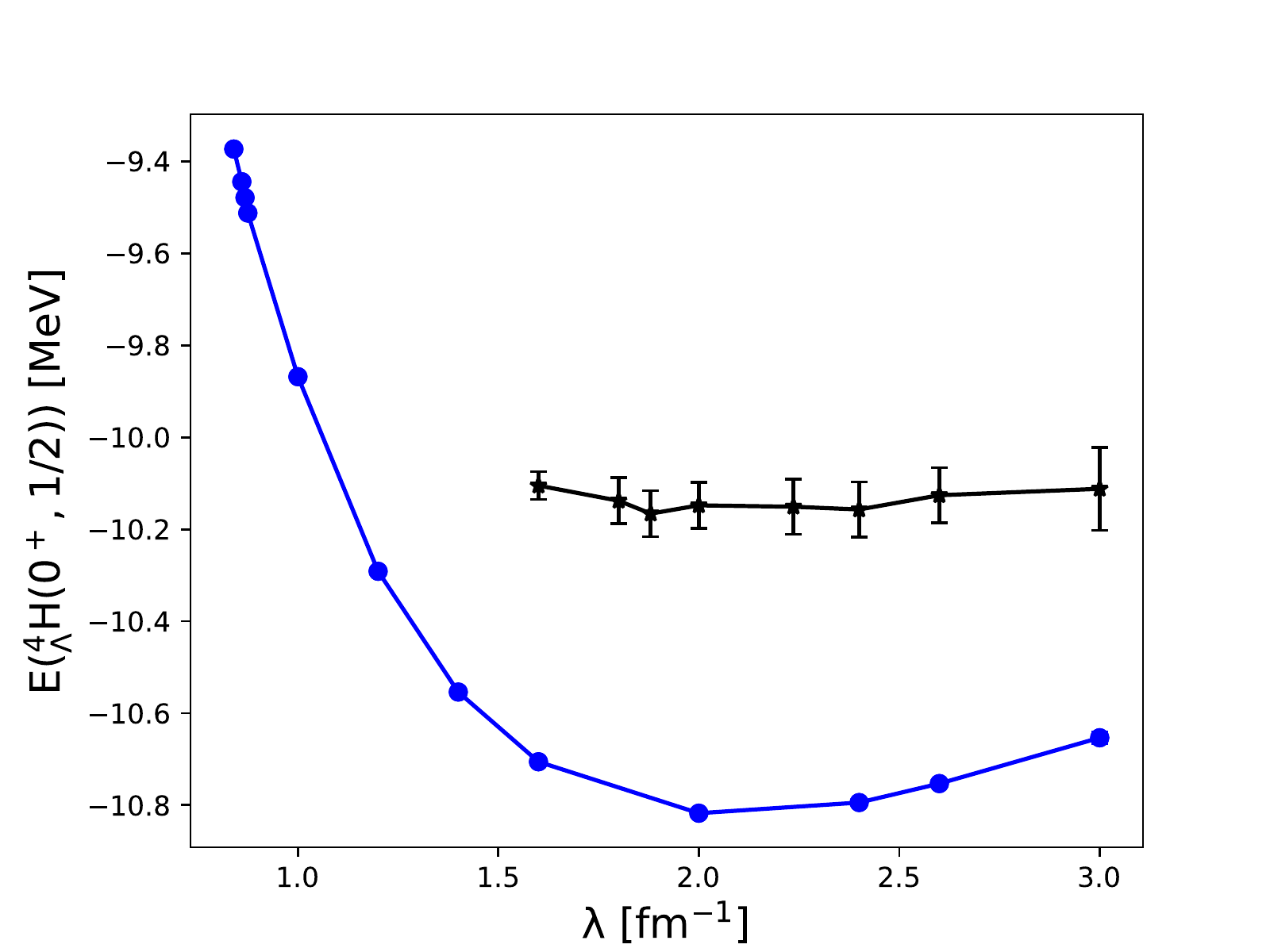}}
              \subfigure[ ]{ \includegraphics[width=0.35\textwidth,trim={0.0cm 0.00cm 0.0cm 0.0cm},clip]{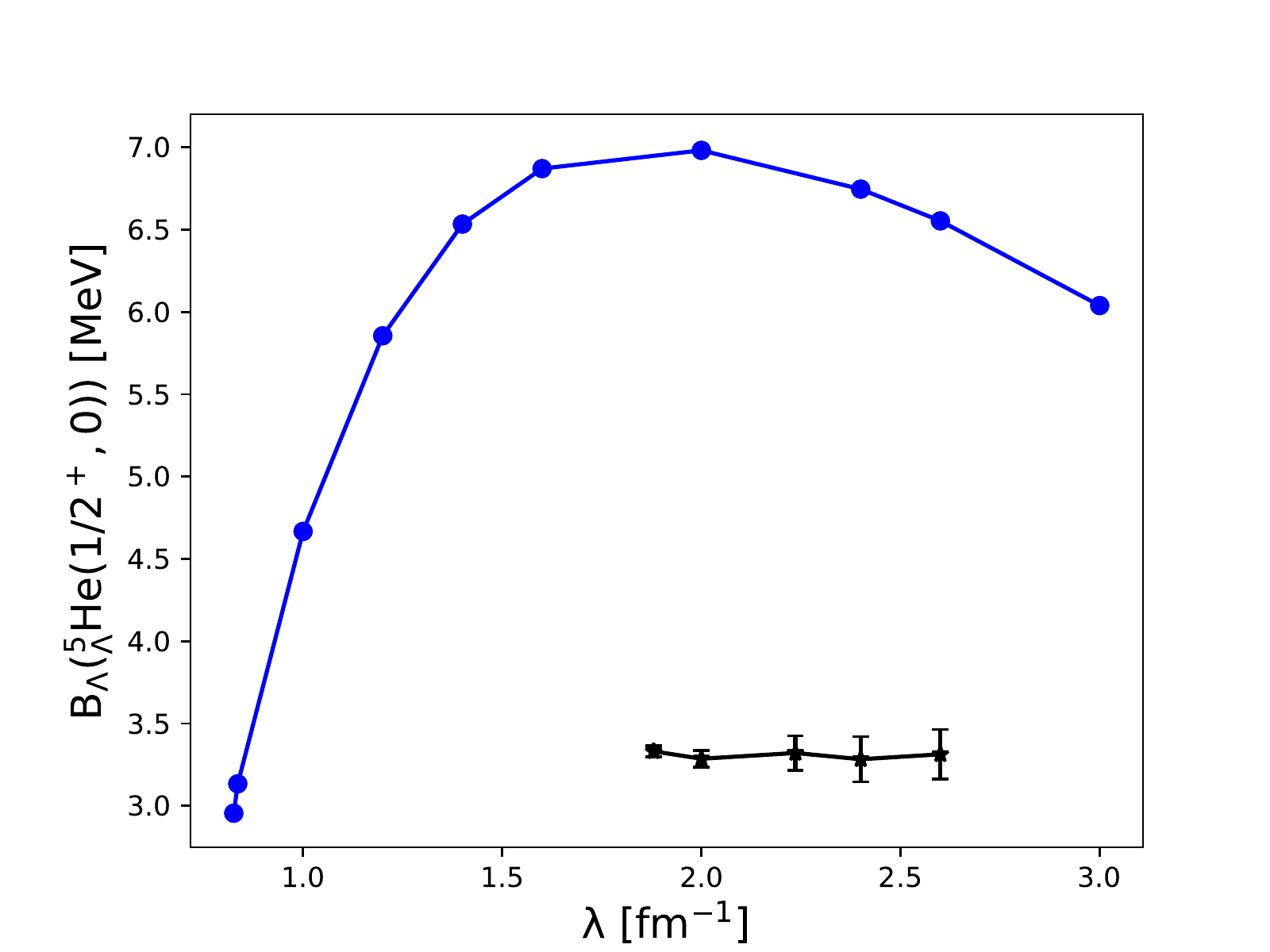}} 
      \end{center}   
      \vspace{-0.8cm}
        \caption{  Dependence of the binding (separation) energies in $A=3-5$ hypernuclei  on the SRG flow
        parameter $\lambda$.  The results (blue circles, panels (a, c,d)) are computed using the  SMS $\mathrm{N^4LO^+450}$ \cite{Reinert:2018} NN potential SRG-evolved to $\lambda_{NN} = 1.6$  fm\textsuperscript{-1} and the  NLO13(500) YN interaction (with CSB included) evolved to a wide range of flow parameter. In panel (b) the strength of the  original NLO13(500) potential  is amplified by a factor of 1.5 so that  $^3_{\Lambda}\mathrm{H}$  is more
        tightly bound.  Black triangles are obtained with  inclusion  of  SRG-induced YNN (panels (a-b)) and of the 3N forces  \cite{Maris:2021} (panels (c-d)). Dashed green lines are binding energies computed by solving  the Faddeev equation for the bare NN and YN interactions.
 }
   \vspace{-0.3cm}
    \label{fig:SRG_dependence}
      \end{figure*}

BB interactions generally contain short-range and tensor correlations that couple low- and high-momentum states,  which make calculations expanded in
 harmonic oscillator (HO) bases such as the NCSM converge slowly.  A very large number of HO functions is required in order to incorporate such strong correlations  and therefore  well-converged results for systems with $A  \ge   5$ are practically computationally inaccessible.  So,  in order soften the interactions we will continuously   evolve the Hamiltonian  using the Similarity Renormalization Group (SRG). The latter can be written in term of a flow equation for  potentials, which, in three-body YNN space,  has following form
   \begin{align}\label{eq:srg}
  \begin{split}
   \frac{dV_s}{ ds}  =  [[   T_{rel}, V_s] ,H_s]; \quad 
    H_s=  T_{rel}    + V_{s}   + \Delta M = T_{rel}   + V^{\mathrm{NN}}_{12,s}  + V^{\mathrm{YN}}_{32,s}  + V^{\mathrm{YN}}_{31,s} + V^{\mathrm{YNN}} _s+ \Delta {M}.
   \end{split}
  \end{align}   
  Here $T_{rel}$ is the intrinsic kinetic energy,   $\Delta M$ is the rest mass difference arising due to particle conversions such as $\Lambda \mathrm{N} - \Sigma \mathrm{N}$, and  
 s is a flow parameter that varies  from 0 to $\infty$ with $s=0$ meaning no evolution or bare interactions. For characterising  SRG-evolved interactions, we will  however use the more intuitive   $\lambda$  parameter, $\lambda= (4\mu_{\mathrm{N}}^2/s)^{1/4}$ with $\mu_{\mathcal{N}}$ being nucleon reduced mass, that provides a measure of the width of the potentials in momentum space \cite{Bogner:2007}. We  solve the flow equation Eq.~(\ref{eq:srg}) in momentum space
 and follow the same approach as done in \cite{Bogner:2007, Hebeler:2012pr} that explicitly separates the flow  equations for two-body NN and YN, and three-body YNN interactions by employing the following relation,
 \begin{align}
 \begin{split}
   \frac{dV_s}{ ds}   =   \frac{dV^{\mathrm{NN}}_{12,s}}{ ds}  +  \frac{dV^{\mathrm{YN}}_{32,s}}{ ds}   +  \frac{dV^{\mathrm{YN}}_{31,s}}{ ds}  +  \frac{dV^{\mathrm{YNN}}_{s}}{ ds}.
 \end{split}
 \end{align}
 In   two-body NN and YN space, the flow equations for $V^{\mathrm{NN}}$ and $V^{\mathrm{YN}}$ can be written as \cite{Le:2020}
 \begin{align}\label{eq:srg2}
 \begin{split}
 \frac{d V^{\mathrm{NN}}_{12}}{ds}  & = [[ t^{\mathrm{NN}}_{12}, V^{\mathrm{NN}}_{12}], \,t^{\mathrm{NN}}_{12} + V_{12}^{\mathrm{NN}} ]\\
\frac{d V^{\mathrm{YN}}_{32}}{ds}  & = [[ t^{\mathrm{YN}}_{32}, V^{\mathrm{YN}}_{32}],\, t^{\mathrm{YN}}_{32} + V_{32}^{\mathrm{YN}}  +\Delta M]; \qquad 
\frac{d V^{\mathrm{YN}}_{31}}{ds}   = [[ t^{\mathrm{YN}}_{31}, V^{\mathrm{YN}}_{31}],\, t^{\mathrm{YN}}_{31} + 
V_{31}^{\mathrm{YN}}+ \Delta M ],
 \end{split}
 \end{align}
 hence, the evolution equation for $V^{\mathrm{YNN}}$ reads,
  \begin{eqnarray}\label{eq:evolVYNN}
&& \hspace{-0.7cm}\frac{d V^{\mathrm{YNN}}}{ds}   =   [[ t^{\mathrm{NN}}_{12}, V^{\mathrm{NN}}_{12}], \,V^{\mathrm{YN}}_{32}  + V^{\mathrm{YN}}_{31} + V^{\mathrm{YNN}}] + [[ t^{\mathrm{YN}}_{32}, \,V^{\mathrm{YN}}_{32}], V^{\mathrm{NN}}_{12} +
V^{\mathrm{YN}}_{31} + V^{\mathrm{YNN}}] \\
  &&    \qquad  +   [[ t^{\mathrm{YN}}_{31},   V^{\mathrm{YN}}_{31}], V^{\mathrm{YN}}_{32}  + V^{\mathrm{NN}}_{12} + V^{\mathrm{YNN}}]  + [[ T_{rel} , V^{\mathrm{YNN}}], T_{rel} + V^{\mathrm{NN}}_{12} + V^{\mathrm{YN}}_{32} + V^{\mathrm{YN}}_{31} +  V^{\mathrm{YNN}} + \Delta M ] \nonumber
  \end{eqnarray}
  Note that now the Kronecker delta functions that appear when embedding two-body interactions in a three-body YNN basis, see the right hand side of Eq.~(\ref{eq:evolVYNN}),  can be evaluated analytically \cite{Bogner:2007, Hebeler:2012pr}.  By projecting Eqs.(\ref{eq:srg2},\ref{eq:evolVYNN})  onto  the corresponding  two-  and three-body  partial-decomposed momentum-space bases, we obtain 
  a system of  coupled ordinary differential equations which will be solved  simultaneously employing the  ARKsover from SUNDIAL.
  In the current work,  we do not include  chiral YNN forces  in the evolution, however, it should be clear from Eq.~(\ref{eq:evolVYNN})  that
the SRG evolution will eventually induce  three-body   YNN   interactions  even if the initial (bare) potential $V^{\mathrm{YNN}}_{s=0}$ is  set to zero, $V^{\mathrm{YNN}}_{s=0}=0$.  Therefore,
omitting those induced higher-body forces, the SRG evolution will  only be  approximately unitary.
Indeed, as it can clearly be seen in Fig.~\ref{fig:SRG_dependence}, the binding energies (square blue symbols) of the $A=3-5$ hypernuclei, 
computed for  SRG-evolved NN and YN interactions, exhibit a strong dependence on the $\lambda$ flow parameter.
  With the inclusion of the induced YNN forces,
we practically reproduce the hypertrition binding energy  $E(^3_{\Lambda}\mathrm{H})$ (dashed green line) that was obtained when solving  the Faddeev equation  for the bare NN and YN potentials, see panel (a) in Fig.~\ref{fig:SRG_dependence}.  Note the large estimated  errors for the NCSM results  because of the extremely small separation energy of  $^3_{\Lambda}\mathrm{H}$.   In panel (b),  we clearly demonstrate  that $E(^3_{\Lambda}\mathrm{H})$  is indeed independent of the SRG  parameter by
increasing the strength of the YN interaction, say,  by a factor of 1.5, so that  the $^3_{\Lambda}\mathrm{H}$ hypernucleus  is  more tightly bound.  Similarly, with both    SRG-induced  YNN and 3N forces included,  the
SRG-dependence of $B_{\Lambda}$  in the   $A=4, 5$ systems is significantly removed, as clearly shown in panels (c-d).  This is an important indication of  likely negligibly  small contributions  of  SRG-induced higher-body forces to the separation energies of these systems. However, for a more quantitative estimate of those contributions, well-converged separation
energies computed for a wide range of flow parameter are  required.

\section{Effect of NLO  YN interactions on $B_{\Lambda}$ in $A=4-7$ hypernuclei}
In our work \cite{Le:2020}, using only  SRG-evolved NN and YN forces,  we have qualitatively studied  predictions of the two practically phase equivalent  NLO13 and NLO19 YN interactions for the separation energies in  the $A=4-7$ systems. Overall, the   NLO19 potential, that is  characterized by a weaker $\Lambda \mathrm{N} - \Sigma \mathrm{N}$ transition potential \cite{Haidenbauer:2020}, yields substantially larger $B_{\Lambda}$ in all systems considered except for $^4_{\Lambda}\mathrm{H}/^4_{\Lambda}\mathrm{He}(0^+)$, see Fig.~10 in Ref.\cite{Le:2020}.  As also  discussed in that work,  such a large  discrepancy between  the NLO13 and NLO19 
 predictions  could be attributed to  possible contributions  of  chiral YNN forces, which according to the underlying power counting will  appear at $\mathrm{N^2LO}$ \cite{Haidenbauer:2022prag,Petschauer:2016}.  
Now, taking into account both chiral and SRG-induced 3N interactions as well as SRG-induced YNN forces, we can study  the actual separation energies
predicted by the two YN  potentials in more detail. These  results together with the experimental $B_{\Lambda}$  for the $A=4-7$ hypernuclei are summarized in Table.~\ref{tab: B_LA=4-7system}.  One sees that 
NLO13 and NLO19  yield  very similar  separation energies for the $^4_{\Lambda}\mathrm{H(0^+)}$ state, with
$B_{\Lambda} = 1.551 \pm 0.007 $ MeV  and $1.514 \pm 0.007$ MeV, respectively, that are  by roughly 0.6 MeV smaller
than the empirical  value of $  B_{\Lambda} = 2.16 \pm 0.08 $ MeV \cite{Juric:1979}.
Likewise, the separation energies for   $^4_{\Lambda}\mathrm{H}(1^+)$ and the ground states in 
$^5_{\Lambda}\mathrm{He}$ and  $^7_{\Lambda}\mathrm{Li}$ computed for NLO13 are $0.823 \pm 0.003$ MeV,   $\,2.22 \pm 0.06$  MeV   and $5.28 \pm 0.68$, respectively. 
 In comparison with  the empirical $B_{\Lambda}$ of $1.07 \pm 0.08$ MeV, 
 $3.12 \pm 0.02$ MeV and $5.58 \pm 0.03$ MeV  \cite{Juric:1979}, respectively,  the NLO13 potential clearly underestimates these systems. In contrast, the  predictions of NLO19 for these three states  are quite close to experiment, only slightly above. 
 For example, the computed separation energies for $^4_{\Lambda}\mathrm{H(1^+)}$ and $^5_{\Lambda}\mathrm{He}$,  $B_{\Lambda}(^4_{\Lambda}\mathrm{H,1^+)}) = 1.514 \pm 0.007$  and $B_{\Lambda}(^5_{\Lambda}\mathrm{He}) = 3.32 \pm 0.03$,  exceed the experiment by less than 0.2 MeV.  Interestingly,  for the ground state of  $^7_{\Lambda}\mathrm{Li}$   we obtained the value of  $B_{\Lambda} = 6.04 \pm 0.30 $  MeV, that    is somewhat  larger than the emulsion 
 separation energy, 
 $B_{\Lambda} = 5.58  \pm 0.03 $ \cite{Juric:1979},  but  it  is in  perfect agreement  with the one deduced from counter experiments, with  $B_{\Lambda} = 5.85 \pm 0.13(10) $ \cite{Agnello:2012}. In conclusion, the discrepancies between $B_{\Lambda}$ predicted by NLO13 and NLO19, and between the computed $B_{\Lambda}$  and  experiment highlight  the importance of chiral YNN forces for properly describing light hypernuclei.

\begin{table}
\vspace{-0.7cm}
   \setlength{\tabcolsep}{0.07cm}
   \renewcommand{\arraystretch}{1.}
\caption{$\Lambda$-separation energies for the $A=4-7$ hypernuclei
computed for the YN potentials NLO13(500) and NLO19(500) including 
the CSB interaction.  The SMS $\mathrm{N^4LO^+}(450)$ and  $\mathrm{N^2LO(450)}$  are used to describe  the  2N and 3N interactions, respectively. All 
the employed potentials are evolved to a SRG  parameter of $\lambda=1.88$ fm\textsuperscript{-1}.  SRG-induced YNN forces are explicitly  taken into account.
All energies are given in MeV.}
\begin{center}
\begin{tabular}{| l |  c c |    c | c |}
\hline
     & \multicolumn{2}{c|}{$ ^4_{\Lambda}\mathrm{H} $}  &  ${^5_{\Lambda}\mathrm{He}}$  &  ${^7_{\Lambda}\mathrm{Li}}$    \\[3pt]
       &  $ 0^+ $    &    $ 1^+$  &   $ {1/2^+} $  &     $(1/2^+,0)$\\   
\hline
\hline
 NLO13(500)    &   $ {1.551 \pm 0.007}  $ &  $ 0.823 \pm 0.003   $   &  $2.22 \pm 0.06$  & $5.28 \pm 0.68 $ \\
NLO19(500)    &   $  {1.514 \pm 0.007}  $ &  $  1.27  \pm 0.009   $   &  $    3.32 \pm 0.03$  & $ {  6.04 \pm 0.30}$\\
\hline
\hline
  \multirow{2}{*}{Exp \cite{Juric:1979,Agnello:2012}   }   &   $2.16    \pm 0.08  $ &  $  1.07  \pm 0.08   $   &  $ 3.12  \pm 0.02$  & $5.85 \pm 0.13(10) $ \\
          &    &     &    & $5.58  \pm 0.03 $ \\
\hline
\end{tabular}
\end{center}
\vspace{-0.4cm}
\label{tab: B_LA=4-7system}
\end{table}

\section{CSB in $A=7,8$ isospin multiplets}
  Being able to accurately compute  hypernuclear wave functions for hypernuclei up to the p-shell, we can now employ those wave functions to explore CSB  effects in the $A=7,8 $ multiplets.  Here, we  employ  the NLO13(500) and NLO19(500) potentials both without and with CSB components \cite{Haidenbauer:2020csb} as YN interactions,  while
    the SMS $\mathrm{N^4LO^+}(450)$ \cite{Reinert:2018}  and $\mathrm{N^2LO}(450)$ \cite{Maris:2021} are  used for  NN and 3N interactions, respectively.  All  the employed potentials are evolved to a SRG  parameter of $\lambda=1.88 $ fm\textsuperscript{-1}.  In addition,  both YNN and 3N forces that are generated during the  evolution are explicitly included  in all  calculations.  
     Let us  further remark that, as one  can also  see in Table.~\ref{tab: B_LA=4-7system},  due to the limited accessible model space,   $B_{\Lambda}$ for systems with $A \ge 7$ are afflicted
with rather large uncertainties, which even exceed the experimentally 
found CSB splittings in those systems.
Therefore, it will be more meaningful to extract the CSB results for the $A=7,\,8$ multiplets perturbatively  based on the  $^7_{\Lambda}\mathrm{Li^{*}( T=1)}$  and $^8_{\Lambda}\mathrm{Li(T=1/2)}$ hypernuclear wave functions and the   $^6\mathrm{Li^{*}(T=1)}$  and $^7\mathrm{Li(T=1/2)}$ nuclear wave functions, respectively.  Note that this perturbative treatment  has been also applied to study  CSB effects  in the $A=4$ systems in Ref.~\cite{Haidenbauer:2020csb}.  The employed 
 wave functions are   computed for the largest computationally accessible model space, namely $\mathcal{N}_{\mathrm{max}} =10 \, (9) $ for $A= 7 \,(8)$, 
and for the 
 optimal HO frequency $\omega_{\mathrm{opt}} =16$~MeV 
 that yields the lowest binding energies. 
 For estimating the uncertainty, we will consider only two-body interactions and  perform calculations using the model space $\mathcal{N}=\mathcal{N}_{\mathrm{max}}, \,\mathcal{N}_{\mathrm{\mathrm{max}}} + 2$, and at the 
 HO frequencies $\omega=\omega_{\mathrm{opt}} \pm 2$~MeV. The variation of the CSB results compared to 
 the values obtained at $\mathcal{N}_{\mathrm{max}}$ and $\omega_{\mathrm{opt}}$ will  be assigned as the uncertainty of our CSB estimations which amount to  30 (50)  keV for  $ A= 7\, (8) $.  Note that the latter is of the same magnitude as that for  the uncertainties of  the CSB results  deduced from experiment,  therefore,   a direct comparison of 
 our CSB predictions with experiment is possible. 
 
 Details of  the  computed CSB splittings for the mirror  hypernuclei  $(^7_{\Lambda}\mathrm{Be}$-$^7_{\Lambda}\mathrm{Li^{*}})$  and 
  $(^8_{\Lambda}\mathrm{Be}$-$^8_{\Lambda}\mathrm{Li})$ 
  are summarized  in Table.~\ref{tab:A7_csb}, where  individual  contributions of the kinetic energy $\Delta T$,  of the NN interaction $\Delta V_{\mathrm{NN}}$ and of  the YN potential  $\Delta V_{\mathrm{YN}}$   to the total CSB are also
  provided. 
Note that the 3N and SRG-induced YNN forces contributions are insignificantly small,  they  are therefore  omitted from the table. Furthermore, in order to  understand the relation between the CSB effect of  the employed  YN potentials and  the $\Delta V_{\mathrm{YN}}$ for larger systems, which
  as shown in   \cite{Haidenbauer:2020csb} contributes dominantly to the CSB splitting in the $A=4$ doublets, we also estimate the $^1S_0$ and $^3S_1$ partial-wave  contributions to $\Delta V_{\mathrm{YN}}$ separately. 
  
 \begin{table}
 \vspace{-0.35cm}
    \setlength{\tabcolsep}{0.1cm}
\renewcommand{\arraystretch}{0.65}
\begin{center}
\begin{tabular}{|l|l|r r|  rrr| r  | }
\hline
\multicolumn{2}{|c|}{} & $\Delta T$ & $\Delta V_\mathrm{NN}$ & \multicolumn{3}{c|}{$\Delta V_\mathrm{YN}$}  & $\Delta B_{\Lambda}$       \\[1pt]
\multicolumn{2}{|c|}{}  &  &  &  $^1S_0$  & $^3S_1$  &  total &  \\
\hline
\hline
 \multirow{4}{*}{$^7_\Lambda$Be-$^7_\Lambda$Li$^*$} & NLO13 &7     & -24          & -1  & 0  & 0             & -17     \\
&  NLO13-CSB & 8    & -24              & -49 & 26   & -24        & -40  \\
\cline{2-8}
&  NLO19 &6     & -40              &-1  & 0        &0     & -34 \\
&  NLO19-CSB &6     & -41              & -43 & 42   & 9        & -35     \\
\cline{2-8}
& Hiyama~\cite{Hiyama:2009ki}  &      &   -70 &    &  &  200 & 150\\
& Gal~\cite{Gal:2015bfa} &  3   & -70  &  &  & 50  &  -17\\
\cline{2-8}
& \,experiment~\cite{Botta:2019has}\, &  &  &  &  &   & $-100 \pm 90 $\\
\hline
\hline
 \multirow{4}{*}{$^8_\Lambda$Be-$^8_\Lambda$Li}
& NLO13          & 12     & 8         &    -2   &   0  & -4      & 16    \\
& NLO13-CSB            & 12     & 7     &   100   &  56    &  159    &  178         \\
\cline{2-8}
& NLO19          & 7      & -11      &    -1     &   0     & -2      & -6    \\
& NLO19-CSB            & 6      & -11       &   62     &  79  &  147    & 143     \\
\cline{2-8}
& Hiyama~\cite{Hiyama:2009ki}  &      & 40   &    &  & & 160\\
& Gal~\cite{Gal:2015bfa}  & 11    & -81  &  & &  119  & 49\\
\cline{2-8}
& experiment \cite{Botta:2016kqd}  &   &    &   &  &  & $40 \pm 60$ \\
\hline
\end{tabular}
\end{center}
\vspace{-0.36cm}
\caption{\label{tab:A7_csb} 
Contributions to CSB in the $A=7$ and $8$ isospin multiplets, 
based on the YN potentials NLO13(500) and NLO19(500) (including 
3N and SRG-induced YNN forces). The results are for the original potentials 
(i.e. without CSB force) and for the scenario CSB1  \cite{Haidenbauer:2020csb}.
 The estimated uncertainties for CSB in $A=7, 8 $ multiplets is 30 and 50 keV, respectively.  All energies are in keV. 
}
\renewcommand{\arraystretch}{1.0}
\end{table}
 One clearly sees that, despite the substantial discrepancies in their predicted $B_{\Lambda}$,
the two YN potentials yield comparable CSB results in the $A=7$ mutiplets. 
 For both cases,  with and without the CSB interactions  included,  the overall CSB effects  are  small 
and  consistent with the experiment  both in magnitude and in sign. It is  further observed that
 the $^1S_0$ and $^3S_1$ 
 states contribute with opposite signs to the total 
 $\Delta V_{\mathrm{YN}}$, like that for the $1^+$ states in the $A=4$ systems,  which in turns leads to a small total CSB in the  $A=7$ isotriplet. 
 Surprisingly,    our prediction for 
$^7_{\Lambda}\mathrm{Be}$-$^7_{\Lambda}\mathrm{Li^{*}}$ for
the  original NLO13 potential  (i.e. without CSB components),  with
$\Delta B_{\Lambda}(\mathrm{NLO13}) = -17$~keV,
is  identical to the value estimated by Gal~\cite{Gal:2015bfa},  for which  a simple shell-model approach in
combination with effective  potentials were employed. However, the individual contributions
$\Delta T$, $\Delta V_{\mathrm{NN}}$, and $\Delta V_{\mathrm{YN}}$ differ substantially. 
For example, the original NLO13    yields a vanishing 
$\Delta V_{\mathrm{YN}}$,  
whereas in Gal's calculation this contribution amounts 
to $50$~keV.  That value is also of opposite sign and larger than our 
estimated $\Delta V_{\mathrm{YN}}$  for the actual CSB  potential. 
In addition,  our NCSM estimate for $\Delta V_{\mathrm{NN}}$ (that also includes the Coulomb effect)
is visibly smaller than the corresponding one used in Gal's calculation. Note that the latter is not computed consistently but rather
 taken from a cluster-model 
study of Hiyama \etal~\cite{Hiyama:2009ki}. 

CSB results for  $^8_{\Lambda}\mathrm{Be}$-$^8_{\Lambda}\mathrm{Li}$ are 
listed at the lower end of Table~\ref{tab:A7_csb}.  
Without  the CSB components,   the two NLO13 and NLO19  potentials  predict
 a negligibly small CSB 
 for $^8_{\Lambda}\mathrm{Be}$-$^8_{\Lambda}\mathrm{Li}$, with
 $\Delta B_{\Lambda} = 16 \pm 50$~keV and $-6 \pm 50$~keV, respectively, which are  well in line with the separation energies difference  of  $40 \pm 60$~keV \cite{Botta:2016kqd}  deduced from  emulsion data. A similarly small   $ \Delta B_{\Lambda}$ was also predicted  by Gal in \cite{Gal:2015bfa}. However, in contrast to the rather small $\Delta V_{\mathrm{NN}}$ contribution in our calculation, e.g. $\Delta V_{\mathrm{NN}} =-11$~keV
 for NLO19, Gal assigned a significantly larger value to  $\Delta V_{\mathrm{NN}}$,  with $\Delta V_{\mathrm{NN}} = -81$~keV that was  taken from a shell model calculation by Millener \cite{Gal:2015bfa}. Interestingly, 
 with the CSB potentials included, both NLO13 and NLO19   yield rather sizable CSB results,  
 $\Delta B_{\Lambda} =177 \pm 50$~keV and  
 $143 \pm 50$~keV, respectively. In this case, the $^1S_0$ and $^3S_1$ partial-wave contributions are  also large, and   importantly,   of the same sign, adding up to a 
 pronounced total CSB. We found that this  exactly resembles the situation 
 for the $0^+$ states of the $A=4$ mirror hypernuclei. One   could therefore   conclude   that a fairly 
 large splitting in the $A=4(0^+)$ states, as  presently suggested by  experiments \cite{Botta:2019has}, 
 implies automatically a likewise significant CSB splitting in 
 ${^8_{\Lambda}\mathrm{Be}}$--${^8_{\Lambda}\mathrm{Li}}$.
 Interestingly, the predictions of NLO13 and NLO19 with CSB interaction are comparable to the value of 
 $\Delta B_{\Lambda} = 160 $~keV obtained in a 
 ($\Lambda$+$\alpha$+$^3$He/$t$) three-body cluster calculation  
 by Hiyama \etal~\cite{Hiyama:2009ki}. 
 However, it should be stressed  that the
 phenomenological CSB YN interaction employed  in that work
 was fitted to 
 an outdated CSB splittings for the $A=4 $ doublets, namely  
 $\Delta B_{\Lambda}(0^+) = 350 \pm 60$~keV and $\Delta B_{\Lambda}(1^+) = 240 \pm 60$~keV. 
 Also it was estimated there that the Coulomb interaction contributes about -80~keV to $\Delta B_{\Lambda}$,  which is substantially larger than our estimate of roughly 10 keV.

\section{Results for $A=4-7$ $\Xi$ hypernuclei}
In this section we  discuss  the results  for $A=4-7$ $\Xi$ hypernuclei   based on our recent study in Ref.~\cite{Le:2021XN}. Here all the calculations have been computed using   the SMS $\mathrm{N^4LO^+}(450)$  NN  \cite{Reinert:2018} and  NLO(500)  $\Xi$N \cite{Haidenbauer:2018gvg}     interactions.  The original $\Xi $N potential  includes coupling to all  other baryon-baryon channels in the
$S=-2$ sector such as $\Lambda \Lambda, \Lambda\Sigma$ and $\Sigma \Sigma$. 
 However,  in practice, the coupling  $\Lambda\Lambda - \Xi$N was omitted here and its contribution  is  effectively incorporated by appropriately   re-adjusting the strength  of the $V_{\Xi\mathrm{N} - \Xi \mathrm{N}}$  potential  \cite{Le:2021XN}. This  facilitates  a proper   convergence of the 
energy calculations to the lowest lying $\Xi$ state.  Furthermore, the  employed  NN  and $\Xi$N potentials are  evolved to a SRG  flow parameter of $\lambda_{\mathrm{NN}} = \lambda_{\Xi \mathrm{N}} =1.6$ fm\textsuperscript{-1}.  For simplicity, the electromagnetic NN interaction \cite{Wiringa:1994wb} as well as the  Coulomb point-like  interaction between $\Xi^-$ and a proton are only included after the evolution. And,  SRG-induced higher-body forces are not included in that study.  As a result, we observe a moderate dependence of  the $\Xi$ separation energies $B_{\Xi}$  on the $\lambda_{\Xi\mathrm{N}}$ parameter  \cite{Le:2021XN}. However, the overall SRG dependence  of $B_{\Xi}$ is relatively small as compared   to the   actual separation energies, and  therefore,  it should not affect our conclusion on the existence of the found bound state \cite{Le:2021XN}.

 The computed   separation energies $B_{\Xi}$ of  $A=4-7$ $\Xi$  hypernuclei together with their  perturbatively  estimated decay width $\Gamma$ are provided in  Table~\ref{tab: A=4-7system} .    It is  estimated that  the $\Xi^- p$ Coulomb interaction  contributes only moderately to the total binding energies of the NNN$\Xi$, $^5_{\Xi}\mathrm{H}$ and  $^7_{\Xi}\mathrm{H}$ systems,  about  200, 600, and  400 keV, respectively. 
  \begin{SCtable}
  \vspace{-0.2cm}
    \renewcommand{\arraystretch}{1.3}
   \setlength{\tabcolsep}{0.22cm}
\begin{tabular}{|c|c c |}
\cline{1-3}
   &  $B_{\Xi}  $ [MeV]  &  $ \Gamma$   [MeV]   \\ 
   \hline
   \hline
$ ^4_{\Xi}\mathrm{H}  (1^+,0) $   &  $0.48 \pm 0.01 $ &   0.74\\
$ ^4_{\Xi}\mathrm{n}  (0^+,1) $ &  $0.71 \pm 0.08$ & 0.2\\
$ ^4_{\Xi}\mathrm{n} (1^+,1) $    & $ 0.64 \pm 0.11$ & 0.01\\ 
$ ^4_{\Xi}\mathrm{H} (0^+,0) $    & - &  - \\ \hhline{  = = =}
$^5_{\Xi}\mathrm{H}(\frac{1}{2}^+,\frac{1}{2})$ &   $2.16 \pm 0.10$ &  0.19  \\ \hhline{  = = =}
$^7_{\Xi}\mathrm{H}(\frac{1}{2}^+,\frac{3}{2})$ &   $3.50 \pm 0.39$ &  0.2  \\ 
  \cline{1-3}
\end{tabular}
\hspace{0.5cm}
\caption{  $  \Xi$ separation energies $B_{\Xi}$ and  estimated decay widths $\Gamma$ for $A=4-7$ $\Xi$ hypernuclei.   All calculations are based on  the NLO(500) YY-$\Xi$N interaction  and the  SMS 
N\textsuperscript{4}LO+(450) NN potential. Both interactions are SRG-evolved to a 
flow parameter of $\lambda_{\mathrm{NN}} = \lambda_{\Xi \mathrm{N}} 
=1.6$~fm\textsuperscript{-1}. The values of   
$B_{\Xi}$ in NNN$\Xi$, $^5_{\Xi}\mathrm{H}$ and $^7_{\Xi}\mathrm{H}$
  are  measured with respect to the binding energies of the core nuclei $^3\mathrm{H}$, $^4{\mathrm{He}}$ and $^6{\mathrm{He}}$, respectively.  }
\label{tab: A=4-7system}
\end{SCtable}
   Thus, the binding of these systems should  be predominantly due to the strong $\Xi$N interaction.  Overall, we found
 three  weakly bound states $(1^+,0)$, $(0^+,1)$ and $(1^+,1)$ in NNN$\Xi$, with quite 
 similar $B_{\Xi}$'s but substantially different decay widths.   Interestingly, our 
  result for  the  ${\rm NNN}\Xi(1^+,0)$ state, $B_{\Xi} = 0.48 \pm 0.01$ MeV,  is close to
  the value of $0.36 \pm 0.16 $ MeV obtained with the  HAL QCD potential \cite{Hiyama:2019kpw}. The latter   however does not support
  the binding   of the two  $(0^+,1)$ and $(1^+,1)$ states.  In addition, there are  
    substantial  differences between our 
    separation energies  $B_{\Xi}({\rm NNN}\Xi)$ and the results 
    for the ESC08c potential \cite{Nagels:2015dia}  also reported in  \cite{Hiyama:2019kpw}. Note that   this version of the Nijmegen potential predicts 
    an unrealistically large strength  for the $\Xi $N interaction in the $^{33}S_{1}$  state that even results in a  bound  $\Xi$N state.

  Our computed  separation energy and decay width  for  $^5_{\Xi}\mathrm{H}$  are   $B_{\Xi}=2.16 \pm 0.1$~MeV  and    
   $\Gamma =0.19$  MeV.  
    Surprisingly, these values agree roughly with the estimations by 
    Myint and Akaishi \cite{Myint:1994qv} of $1.7$~MeV and $0.2$~MeV, 
    respectively. However,  in contrast 
    to our finding that  $^5_{\Xi}\mathrm{H}$ is bound primarily due to the strong $\Xi$N interaction, the authors in \cite{Myint:1994qv} estimated that most of the $1.7$ MeV binding energy comes from the   $^4\mathrm{He}$-$\Xi^-$ Coulomb  interaction.  A quite 
    similar separation energy, $B_{\Xi}(^5_{\Xi}\mathrm{H})=$   2.0~MeV, is also reported    in the work by  Friedman and Gal,  where  an optical potential   is employed \cite{Friedman:2021rhu}. 
    But, as discussed in \cite{Le:2021XN},  such an  agreement could be more or less accidental.
        
   For  the  $^7_{\Xi}\mathrm{H}$ $(\frac{1}{2}^+, \frac{3}{2})$ system,  the chiral $\Xi$N potential yields a  separation energy
   of  $B_{\Xi}=3.50 \pm 0.39$~MeV  and a narrow   decay width, with   $\Gamma= 0.2 $ MeV.  The  computed  $B_{\Xi}$
 is only slightly larger than the value  of 3.15~MeV reported by Fujioka \etal 
 \cite{Fujioka:2021S}  when the  HAL QCD potentials is employed in  combination with 
a four-body ($\alpha nn \Xi$) cluster model
 \cite{Hiyama:2008fq}. With an    older $S=-2$ potential from the Nijmegen group, the authors however obtained 
  somewhat 
 smaller binding energy,  namely $B_{\Xi} = 1.8 $ MeV \cite{Hiyama:2008fq}. 
 
      Let us finally  remark that
 in  Ref.~\cite{Le:2021XN}  we  have also explored the relation between  properties of the employed chiral $\Xi $N interaction and the binding of the $A=4-7$ $\Xi$ systems. We shown that the $\Xi$N attraction in the $^{33}S_{1}$ state 
is essential for the binding of all the systems considered, which contributes  to more than $50 \%$ of the $\Xi$N potential expectation value $\langle V_{\Xi\mathrm{N}}\rangle$.

 \section{Conclusion}
  In this work, we  throughly studied  single- and double-strangeness hypernuclei with $A=4-8$ based on the Jacobi NCSM and in combination with the NN (3N), YN and $\Xi$N  interactions consistently  derived within the chiral EFT approach. In order to speed up the convergence of the NCSM with respect to model space, all the employed two-body and 3N potentials are transformed using an SRG evolution.
  We  explicitly showed that,  with the inclusion of the SRG-induced three-body forces (3N and YNN), the strong dependence of  $B_{\Lambda}$
  on  SRG flow parameter is removed for $^3_{\Lambda}\mathrm{H}$ and significantly reduced for the $A=4, 5$ systems. 
  After that, we studied the predictions of the two practically phase-equivalent   NLO13(500) and NLO19(500) YN interactions   for the separation energies in the $A=4-7$ hypernuclei.  Both potentials underestimate the ground state of $^4_{\Lambda}\mathrm{H(0^+)}$.
   The NLO19 potential yields $B_{\Lambda}$ for $^4_{\Lambda}\mathrm{H}(1^+)$ and the ground states in $^5_{\Lambda}\mathrm{He}$ and $^7_{\Lambda}\mathrm{Li}$ that are comparable with the experiments (only slightly above), whereas  NLO13 substantially
    underestimates these systems.  We also investigated CSB splittings  in the $A= 7,\,8$  multiplets using the two NLO YN  potentials that also include  the leading  CSB interaction in the $\Lambda $N channel  derived in Ref.~\cite{Haidenbauer:2020csb}.  Overall, the predicted CSB for the $A=7$ systems are small and agree with experiment while the results for the  $A=8$ doublet 
  are somewhat larger than the available empirical value deduced based   on emulsion data.
  Finally, we discussed our recents results for $\Xi$ hypernuclei with $A=4-7$ based on the NLO(500) $\Xi$N potential. We 
  found three loosely bound states $(1^+,0)$, $(0^+,1)$  and $(1^+,1)$ for the NNN$\Xi$ system. The
    $^5_{\Xi}\mathrm{H}$ and $^7_{\Xi}\mathrm{H}$ $(\frac{1}{2}^+, \frac{3}{2})$  hypernuclei are more tightly bound.   The decay widths of these  $\Xi$ bound states are expected to be very  small which could allow them to   be observed in experiment.  
   
 \vspace{0.3cm}
\noindent{\textbf{Acknowledgements}}:\small{ I am very grateful to   Ulf-G. Mei{\ss}ner, Johann Haidenbauer and Andreas Nogga for the fruitful collaboration,  for proofreading and providing valuable suggestions to improve  the manuscript. 
This work is supported   by the  DFG and NSFC  through the funds provided to the Sino-German Collaborative
Research Center TRR110 ``Symmetries and the Emergence of Structure in QCD''.  Numerical calculations
have been performed on JURECA and 
JURECA booster of the JSC, J\"ulich, Germany.   }

\end{document}